\documentclass[twocolumn,pre,aps,superscriptaddress]{revtex4-1}
\usepackage{times}
\usepackage{mathptm}
\usepackage{graphicx,epsfig}
\usepackage{amsmath}

\usepackage{graphicx}

\usepackage[scaled]{helvet} 
\usepackage{courier} 
\normalfont
\usepackage[T1]{fontenc}

\begin{document}

\title{Unwinding
and rewinding the nucleosome inner turn:
Force dependence of the kinetic rate constants}
\author{S. G. J. Mochrie}
\email{simon.mochrie@yale.edu}
\thanks{Corresponding author;  phone: 203 436-4809}
\affiliation{Integrated Graduate Program in Physical and Engineering Biology,
Yale University, New Haven, Connecticut 06520}
\affiliation{Department of Applied Physics, Yale
University, New Haven, Connecticut 06520}
\affiliation{Department of Physics, Yale
University, New Haven, Connecticut 06520}
\author {A. H. Mack}
\affiliation{Integrated Graduate Program in Physical and Engineering Biology,
Yale University, New Haven, Connecticut 06520}
\affiliation{Department of Applied Physics, Yale
University, New Haven, Connecticut 06520}
 \author{D. J. Schlingman}
\affiliation{Integrated Graduate Program in Physical and Engineering Biology,
Yale University, New Haven, Connecticut 06520}
\affiliation{Department of Molecular Biophysics and Biochemistry,
Yale University, New Haven,
Connecticut 06520}
 \author{R.  Collins}
 \affiliation{Department of Molecular Biophysics and Biochemistry,
Yale University, New Haven,
Connecticut 06520} 
 \author{M.  Kamenetska}
 \affiliation{Department of Physics, Yale
University, New Haven, Connecticut 06520}
 \affiliation{Department of Molecular Biophysics and Biochemistry,
Yale University, New Haven,
Connecticut 06520} 
 \author{L.  Regan}
\affiliation{Integrated Graduate Program in Physical and Engineering Biology,
Yale University, New Haven, Connecticut 06520}
\affiliation{Department of Molecular Biophysics and Biochemistry,
Yale University, New Haven,
Connecticut 06520} 
\affiliation{Department of Chemistry,
Yale University, New Haven,
Connecticut 06520}


\date{\today}


\begin{abstract}
A simple model for the
force-dependent unwinding and rewinding rates of the nucleosome inner turn is
constructed and  quantitatively compared to the results of recent measurements
[A. H. Mack {\em et al.}, J. Mol. Biol.  {\bf 423}, 687 (2012)].
First, a coarse-grained model for the histone-DNA free energy landscape
that incorporates both an elastic
free energy barrier and specific histone-DNA bonds is developed.
Next, a theoretical expression for the rate of transitions across a piecewise
linear free energy landscape with multiple minima and maxima is presented.
Then, the model free energy landscape, approximated as a piecewise linear
function, and the  theoretical
expression for the transition rates are combined to construct a model for the
force-dependent unwinding and re-winding rates of the nucleosome inner turn.
Least-mean-squares fitting of the model rates to the rates observed in recent
experiments rates demonstrates that this model is able to well describe the 
force-dependent unwinding and rewinding rates of the nucleosome inner turn,
observed in the recent experiments,
except at the highest forces studied, where an additional {\em ad hoc} term is required to describe the data,
which may be interpreted as an indication of an alternate high-force nucleosome disassembly pathway, that
bypasses simple unwinding.
The good agreement  between the measurements and the model at lower forces
demonstrates that both specific histone-DNA contacts and an elastic free energy barrier
play essential roles for nucleosome winding and unwinding, and quantifies their relative contributions.


\end{abstract}

\maketitle

\section{Introduction}
\label{INTRODUCTION}
To fit into a nucleus, eukaryotic DNA is assembled by histones and other proteins into a hierarchy of chromatin structures, and ultimately chromosomes.
The fundamental organizational unit of chromatin is the nucleosome in which
146 or 147 bp of DNA are wound around a protein complex comprised
of two copies each of the core histones, H2A, H2B, H3 and H4.
The  core histones are highly conserved across Eukarya, and all have a
similar general structure consisting of a central globular domain, an unstructured N-terminal tail,
and an unstructured C-terminal tail. The structure of the canonical nucleosome is known
in atomic detail as a result of x-ray crystallographic studies \cite{Luger1997,Davey2002,Chua2012}.

By blocking access to promotor DNA, nucleosomes generally repress gene expression in eukaryotes.
For highly transcribed genes, the depletion of nucleosomes from
the promoter DNA is a key feature and is believed to be a prerequisite
for the recruitment of TATA-binding protein and RNA polymerase
\cite{Boeger2003,Adkins2007,Kim2008}.
Nucleosome eviction from the promotor can be
accomplished by a number of ATP-dependent chromatin
``remodeling'' enzymes \cite{Saha2006}, such as SWI/SNF,
which exert force to displace the nucleosomes.
It follows that the behavior of nucleosomes
under tension, and their mechanical properties more generally, are
highly relevant for eukaryotic gene expression.
In addition, chromatin is subjected to significant forces during cell division.

\begin{figure}[t!]
	\begin{center}
		\includegraphics[width=3.2in]{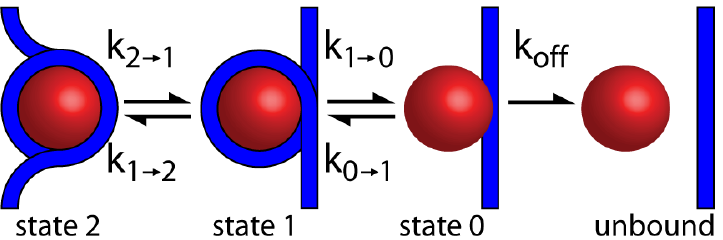}
	\caption{ (Color onine)
The four microstates for 
nucleosome unwinding/rewinding
proposed in Ref. \onlinecite{Brower2002}.
The histone octamer is depicted as a circle.
DNA is depicted as a line.
For  state 2, the nucleosome is fully wrapped by nearly two turns of DNA.
For state 1, the outer turn is unwrapped, but the inner turn is wrapped.
For state 0, both the outer and inner turns are unwrapped,
but the histone octamer remains bound to the DNA.
Finally, the histones may be unbound from the DNA.
}
\label{Fig1}
	\end{center}
\end{figure}

Motivated by these considerations,
several groups have sought to characterize and understand the forced unwinding of
nucleosomes using optical or magnetic tweezers
\cite{Cui2000,Bennink2001,Brower2002,Pope2002,KulicSpools,Brower2005,Gemmen2005,Pope2005,Mihardja2006,Kruithof2009b,Kruithof2009,Hall2009,Simon2011,Molla2012}.
The  accepted model for the nucleosome winding/unwinding pathway, first
proposed by Brower-Toland {\em et al.}
 \cite{Brower2002},
 is illustrated in Fig.~\ref{Fig1}.
The model  envisions four distinct nucleosome states,
designated state  2, state 1, state 0, and unbound.
State 2 is the canonical nucleosome, in which the histone octamer
is wrapped by nearly two turns of DNA.
For state 1, the inner turn remains wrapped, but the outer turn is unwrapped.
For state 0, both the outer and inner turns are unwrapped,
but the histone octamer remains bound to the DNA.
Finally, for  the unbound state, the histone octamer is dissociated from the DNA.

The purpose of this paper is to present a physics-based model for the
force-dependent unwinding and rewinding rates of the nucleosome inner turn,
 that well describes recent experimentally measured unwinding and rewinding rates,
shown in Fig.~\ref{RATES} \cite{Mack2012a}.
To this end, we first develop a simple coarse-grained model for the histone-DNA free energy landscape
that incorporates both the elastic
free energy barrier described in Refs. \onlinecite{KulicSpools,Mihardja2006,Sudhanshu2011}
and specific histone-DNA bonds, as indicated in Ref.  \onlinecite{Hall2009}.
Secondly, we  present a theoretical expression for the rate of transitions across a piecewise
linear free energy landscape with multiple minima and maxima.
Thirdly, we combine our model free energy landscape, approximated as a piecewise linear
function, and our theoretical
expression for the transition rates across a piecewise-linear landscape
to construct a  model for the
force-dependent unwinding and re-winding rates of the nucleosome inner turn.
Finally, we demonstrate by least-mean-squares fitting that this model provides a good description of  the observed
force-dependence of the unwinding and rewinding
rates of the nucleosome inner turn,
as shown as the solid and dashed lines in Fig.~\ref{RATES}.
At high forces, an additional {\em ad hoc} term is required to describe the data,
which we interpret as an indication of an alternate high-force nucleosome disassembly pathway, that
bypasses simple unwinding.
Nevertheless, the good agreement  between our measurements and our model at lower forces
indicates that both specific histone-DNA contacts and an elastic free energy barrier
play essential roles for nucleosome winding and unwinding, and it quantifies their relative contributions.

\section{Experimental review}
Using optical tweezers,
Brower-Toland {\em et al.} \cite{Brower2002} studied the forced unwinding of nucleosome arrays,
assembled from natural avian histones on a
sequence of  DNA containing 17 tandem repeats of the 5S rRNA nucleosome positioning sequence.
They found that under tension  a length of DNA equivalent to about 75 bp per nucleosome
unwinds apparently continuously at lower forces,  corresponding to unwinding of the
nucleosome outer turn of DNA.
At higher forces,  the force-versus-extension curve shows discrete jumps,
each of about 75~bp,
each of which corresponds to unwinding the inner turn of DNA.
Brower-Toland {\em et al.} ascribed the observed gradual unwinding of the nucleosome
outer turn to relatively weak outer-turn histone-DNA interactions, and
the 
high-force unwinding of the nucleosome inner turn --
the transition from state 1 to state 0 --
to the sudden disruption of strong histone-DNA interactions,
located between 35 and 45~base pairs either side of the nucleosome dyad.

Kulic and Schiessel \cite{KulicSpools} proposed an alternative explanation for the sudden
inner turn disruption events, observed by Brower-Toland
{\em et al.}.
They showed that the elasticity of a nucleosomal DNA superhelix under tension
gives rise to a force-dependent free energy barrier between states with different
numbers of turns.
This interpretation found support in subsequent experiments,
which studied the behavior of
avian mononucleosomes under tension, assembled
on DNA containing a single 601 nucleosome positioning sequence
\cite{Mihardja2006,Kruithof2009b,Sudhanshu2011}. 
Consistent with the predictions of
Ref. \onlinecite{KulicSpools}, these experiments reveal
that for tensions near 3~pN, the mononucleosome ``hops'' between states 1 and 2,
indicating the existence of a significant free energy barrier between these states too.

Recently, to further elucidate  the energetics of histone-DNA binding,
as well as to probe the forces experienced by
 a polymerase molecule proceeding through nucleosomal DNA
 during transcription,
Hall {\em et al.} \cite{Hall2009,Forties2011} implemented a novel approach, using
optical tweezers to unzip double-stranded DNA, wound around a nucleosome, into two single-stranded DNA strands.
As the unzipping fork proceeds around the nucleosome,
Hall {\em et al.} found that
it dwells at discrete positions, spaced one from another by about 5~bp,
indicating the existence of free energy barriers to unzipping, spaced by about 5~bp.
Because the dwell times observed were much larger than in the absence of
the nucleosome, Hall {\em et al.} were able to infer that  such barriers are the result of discrete
histone-DNA binding sites, spaced by approximately 5~bp.

In addition to the  5~bp periodicity, Hall {\em et al.} also
identified three broad regions within each of which the dwell time is especially long,
indicating the existence of large free energy barriers to unzipping in these regions.
The largest barrier occurs near the nucleosome dyad (region 2),
but there are two additional regions
in which the barriers are high, 
displaced from the dyad by about  $\pm45$~bp
(region 1 near $-45$~bp and region 3 near +45~bp).
Barriers are largest in region 2, then comes region 1, with
the smallest barriers in region 3.
Hall {\em et al.}'s collected results convincingly demonstrate the existence of  significant
localized histone-DNA interactions,
which should play an essential role in nucleosome unwinding and re-winding,
because unwinding necessarily disrupts these bonds.

More recently still, several of us employed optical tweezers in
a force-clamp mode to determine, for the first time,
the unwinding
and rewinding rates of the nucleosome inner turn at
several fixed  forces ($F$)
\cite{Mack2012a}.
We studied
homogeneous nucleosomes containing recombinant wild-type H2A, H2B, H3,
and either recombinant wild-type H4 (henceforth H4)
or recombinant mutant H4 with arginine 45 mutated to histidine
(henceforth H4-R45H).
which is an example of a so-called
 {\underline S}WI/SNF-{\underline i}{\underline n}dependent
(Sin) mutation
\cite{Sternberg1987,Kruger1995,Wechser,Muth2004,Flaus2004,Kurumizaka}.
Our motivation for measuring these rates was the observation that
the upstream regulatory region ({\em URS1}) of the yeast {\em HO} promoter
contains a strong nucleosome positioning sequence, which
ordinarily requires  SWI/SNF for transcription to occur, but that
in the absence of SWI/SNF,  Sin mutant histones 
restore transcription from the {\em HO} locus to wild-type levels.
Thus, quantification of the differences between wild-type nucleosomes and nucleosomes containing
Sin mutant histones would represent a direct measurement of the
minimal free energy change required for an {\em in vivo} effect,
with broad implications for the mechanisms by which histone modifications
 \cite{Jenuwein2001,Fraga2005b} and histone variations
\cite{Ausio2006,montel2009,henikoff2008}
affect gene expression.
Nucleosomes assembled with H4-R45H also show significantly enhanced thermally-driven nucleosome sliding
compared to nucleosomes containing H4
\cite{Flaus2004}.

\begin{figure*}[t!]
\begin{center}
\includegraphics[width=5.4in,keepaspectratio=true]{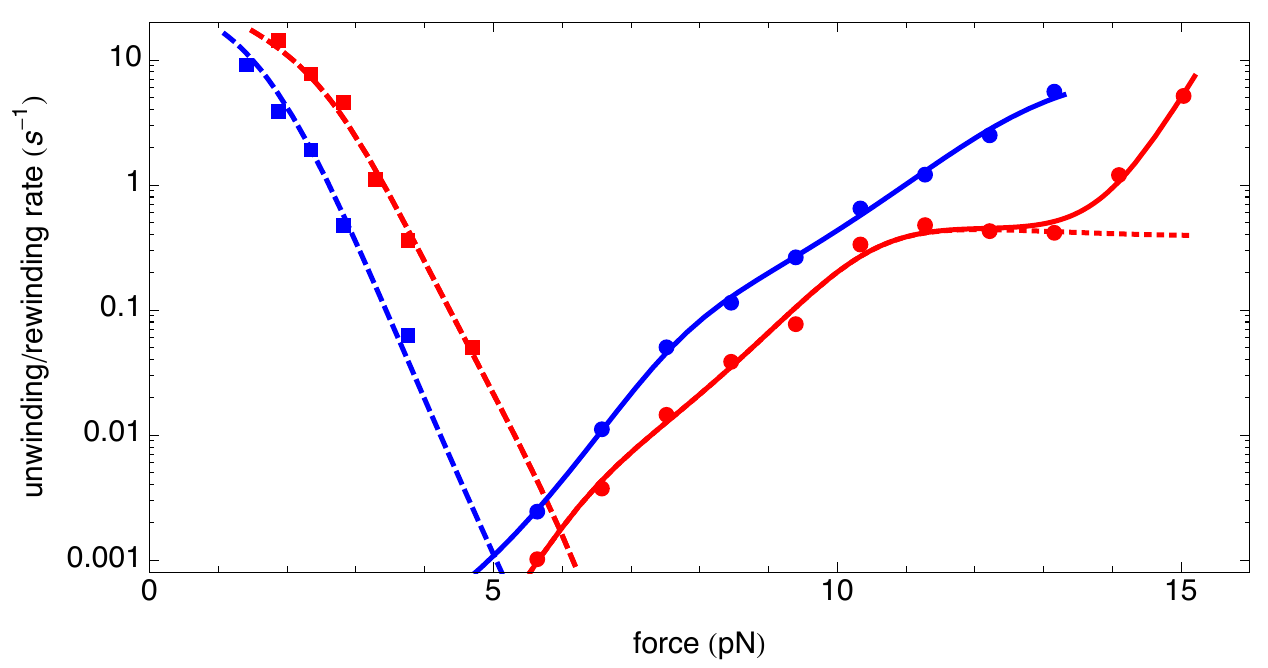}
\end{center}
\caption{ (Color online)
Comparison of the unwinding and rewinding rates of the nucleosome inner turn
for nucleosomes containing H4 (red) or H4-R45H (blue)
as a function of force.
Circles correspond to the unwinding rate ($k_{1 \rightarrow 0}$).
Squares correspond to the re-winding rate ($k_{0 \rightarrow 1}$).
The solid and dashed lines are results of least-mean-squares fits to the model
described in the text.
The upper dashed line and squares correspond to H4.
The lower dashed line and squares  correspond to H4-R45H.
The upper solid line and circles correspond to H4-R45H.
The lower solid line and circles correspond to H4.
The dotted line is the model, but omitting the {\em ad hoc} term,
described in the text.
}
\label{RATES}
\end{figure*}

Our measurements of the unwinding  and rewinding rates of the nucleosome inner turn
 are reproduced in Fig.~\ref{RATES} with the rates displayed on a logarithmic axis,
and the corresponding forces displayed on a linear axis.
The measured unwinding rates span nearly four orders of magnitude, and the
measured rewinding rates more than two orders of magnitude.
At each force tested, nucleosomes containing H4-R45 unwind more rapidly and
re-wind more slowly than nucleosomes containing H4, indicating that the latter are the more stable.
In fact,
using the data of Fig.~\ref{RATES}, we were able to determine the free energy of the nucleosome inner turn.
In brief, 
there are two contributions to the force-dependent Gibbs free energy difference
between state 0 and state 1 ($G_{total}$), namely
the free energy of the nucleosome inner turn ($G_{0 \rightarrow 1}$) and
the difference in the Gibbs free energy of the
DNA, not wound around the nucleosome  \cite{Tinoco2002}.
At the force, $F^*$, at which the unwinding rate and the rewinding rate are equal,
state 1 and state 0 have the
same Gibbs free energy, and $\Delta G_{total}=0$.
It follows that
\begin{equation}
G_{0 \rightarrow 1}
= - F^* d \frac{ 1- \sqrt{ \frac{k_B T}{ F^* L_P }} } { 1- \sqrt{ \frac{k_B T}{4F^* L_P }}   },
\end{equation}
where $d$ is the difference in extension between state 0 and state 1.
By extrapolation of the rates in Fig.~\ref{RATES},
we find $F^*= 6.0 \pm 0.3$~pN  for  nucleosomes containing H4,
and $F^*= 5.0 \pm0.3$~pN for nucleosomes containing  H4-R45H.
 It follows, using  $L_P \simeq 42$~nm,
 and  $d= 23.3\pm0.5$~nm at 5~pN and
$d= 23.6 \pm 0.5$~nm at 6~pN
\cite{Mack2012a}
 that
$G_{0 \rightarrow 1} = -142 \pm 7~$pNnm$ = -34.6\pm1.7 k_B T= -20.8\pm1.0$~kcal/mol
 for nucleosomes containing H4,
and $G_{0 \rightarrow 1} = -117 \pm7$~pNnm $= -28.5\pm1.7 k_B T= -17.2\pm1.0$~kcal/mol
 for nucleosomes containing  H4-R45H.
The difference in the free energy of the nucleosome inner turn for nucleosomes with H4 and H4-R45H 
is therefore $6.1 \pm 2.4 k_B T$ \cite{Mack2012a}.

\begin{figure}[t!]
 {\includegraphics[width=2.86in,keepaspectratio=true]{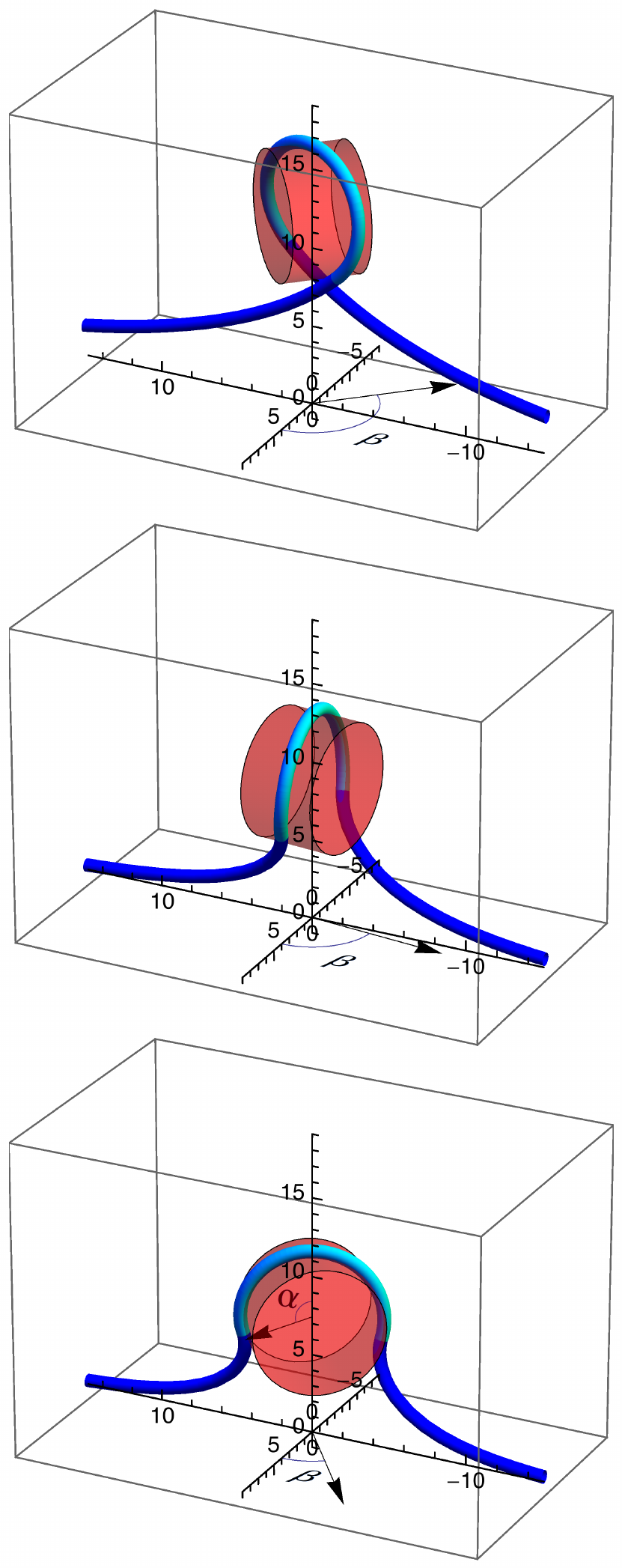}}
\caption{ (Color online)
Elastic-rod models of DNA,  depicted as a
line,  partially wound around a
cylindrical representation of the histone octamer, depicted as a 
partially-transparent
cylinder, at a force of 8~pN for three different winding ($2\alpha$) and rotation
($\beta$) 
angles: 
$2\alpha=4.8$~radians  and  $\beta = 2.33$~radians (top);
$2\alpha=4.18$~radians  and  $\beta = 1.46$~radians (middle); and
$2\alpha=3.7$~radians  and  $\beta = 0.65$~radians (bottom).
DNA in contact with the nucleosome is shown lighter than the DNA not in
contact with the nucleosome.
The angle subtended by this DNA defines the winding angle $2 \alpha$,
as indicated in the bottom panel.
The rotation angle ($\beta$) of the cylinder axis about the vertical axis is indicated
in each panel.
Axes are marked in nanometers.
}
\label{model}
\end{figure}

The usual model for how force affects reaction rates, introduced in Ref. \onlinecite{Bell1978},
predicts an exponential variation of the rates with force, leading to linear curves,
when plotted on semi-logarithmic axes.
By contrast,  the measured unwinding and
rewinding rates, presented in Fig.~\ref{RATES},
show significant curvature. Most notable in this regard is that the unwinding rate of nucleosomes
containing H4  shows a nearly force-independent rate for forces between about 10 and 13~pN.
However, the other three curves also  show changes in slope, albeit less dramatic.
These deviations from the usual behavior motivate  re-consideration of
how to appropriately describe these data.
(Recent elaborations of  Ref. \onlinecite{Bell1978}
predict curvature in semi-logarithmic plots  \cite{PhysRevLett.96.108101}.
However,
the predicted curvature is small, only becoming noticeable for rates extending over 7 or 8 orders of magnitude.
The changes in slope, we observe, are much larger.)

\begin{figure}[t!]
      {\includegraphics[width=3.2in,keepaspectratio=true]{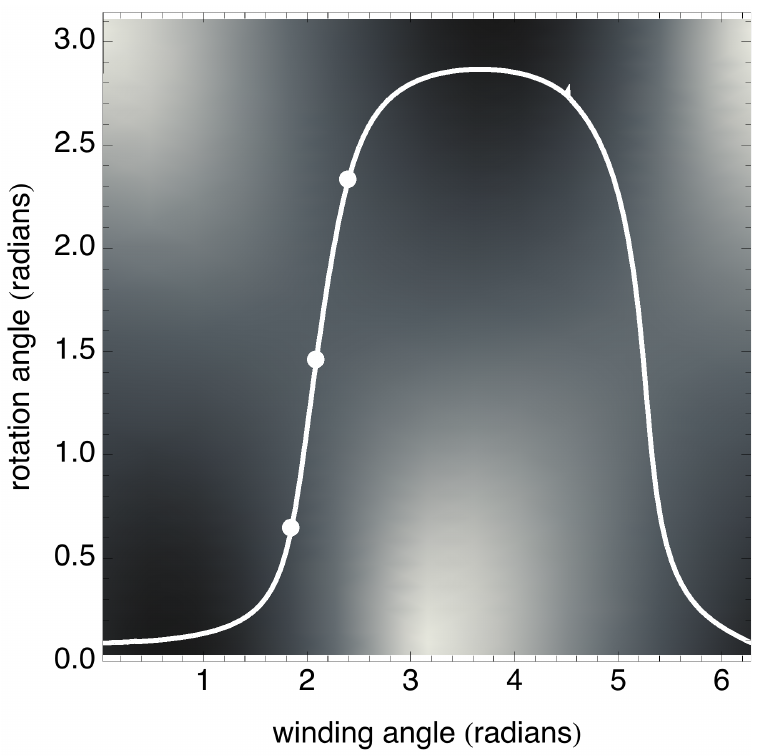}}
\caption{
Grey scale representation of the 
periodic parts of the elastic free energy at a fixed force of $8$~pN, {\em i.e.} the final two terms in EQ. \ref{GIBBS_ELASTIC},
for which white corresponds to the highest elastic energy and black the lowest.
The white line shows the rotation angle, $\beta$, versus winding angle, $\alpha$,  for $F=8$~pN,
according to EQ. \ref{IMPLICIT}.
The points correspond to the winding angles and rotation angles for the
configurations shown in Fig.~\ref{model}.
}
\label{LANDSCAPE}
\end{figure}

 \begin{figure}[b!]
\begin{center}
{\includegraphics[width=3.2in,keepaspectratio=true]{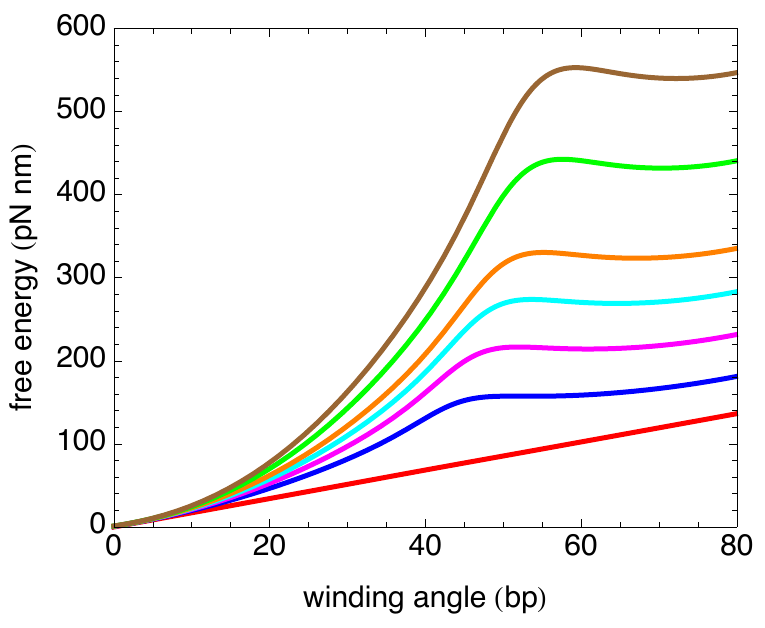}}
\end{center}
\caption{
(Color online) Elastic free energy landscape for the nucleosome inner turn
at several forces, from bottom to top:
0,
2, 
4, 
6,
8, 
12, 
 and
16~pN. 
}
\label{LANDSCAPE2}
\end{figure}

\section{Model free energy landscape of a nucleosome under tension}
\subsection{Elastic free energy of a DNA superhelix under tension}
\label{HELIX_UNDER_TENSION}
In this section, we present a model for the force-dependent elastic free energy of
nucleosomal DNA, treated as an  elastic rod, following Ref. \cite{KulicSpools}.
Our calculations are carried out for a single nucleosome, flanked by infinitely-long DNA arms.
In comparison,  the experiments, that we seek to describe \cite{Mack2012a},
are carried out on 12-nucleosome arrays, held at forces varying from $F=1.5$ to 15~pN,
in which the nucleosomes are in state 1 and are
separated from each other by about 40~nm. The characteristic
length for elastic deformations of the DNA is given by $\lambda=\sqrt{\kappa/F}$,
where $\kappa$ is the bending modulus of the DNA.
Thus, $\lambda$ varies between 11~nm to 3.4~nm, depending on the force.
Since $\lambda$ is always many times less than the separation between nucleosomes,
our treatment, based on isolated nucleosomes,  seems reasonable.
In addition, we assume torsionally unconstrained DNA, which also corresponds to the
experimental situation that we are seeking to describe, because in optical tweezers experiments,
the bead, and the DNA that it is attached to, are free to rotate.
Because we treat DNA as an elastic rod,  our calculation does not treat the role of different
DNA sequences. Our calculations also assume that linker histone H1 is not present, also corresponding
to the experimental situation \cite{Mack2012a}.

Illustrated in Fig. \ref{model} are three model configurations of nucleosomal DNA.
In each case, the DNA is shown as a  rod,
partially wound around a histone octamer, represented as a partially-transparent cylinder.
%
To model the DNA's elastic energy, we start with the expression, adapted from
Ref. \onlinecite{KulicSpools},
for the elastic Gibbs free energy ($\Delta G_{elastic}$) of DNA that is wound into a superhelix,
attached to DNA ``arms'', and
subjected to a tension, $F$:
\begin{widetext}
\begin{eqnarray}
\Delta G_{elastic}(F)
=&&
 \frac{\kappa \alpha}{R} +(2\alpha R-L_C)F \left ( 1- \frac{k_BT}{\sqrt{\kappa F}} \right )
-2 F R \left(\cos \beta \sin \alpha -\frac{H \alpha  \sin \beta }{R}\right)
\nonumber \\
&&+\sqrt{\kappa F } \left(8-4 \sqrt{2+\frac{2 R \cos \alpha \cos \beta }{\sqrt{H^2+R^2}}+\frac{2H \sin \beta }{\sqrt{H^2+R^2}}}\right).
\label{GIBBS_ELASTIC}
\end{eqnarray}
\end{widetext}
In Eq. \ref{GIBBS_ELASTIC},
$\kappa$ is the bending modulus of DNA, $L_C$ is its total contour length,
$R$ is the superhelical radius, $2 \pi H$ is the superhelical pitch,
$2 \alpha$ is the winding angle of the DNA about the histone octamer,
shown green in Fig. \ref{model}, and $\beta$ is the nucleosome rotation
angle.
For a nucleosomal superhelix, we take $R=4.18$~nm, $2 \pi H = 2.39$~nm ($H=0.38$~nm) \cite{Luger1997},
and
$\kappa = 172$~pN nm$^{2}$, corresponding to a DNA persistence length of $L_P = {\kappa}/{ (k_B T)} = 42$~nm.
We routinely find that a value of $L_P = 42$~nm well describes the
force-versus-extension of dsDNA at  the solution conditions of our experiments.

The periodic terms in EQ. \ref{GIBBS_ELASTIC} give rise to a force-dependent barrier
between configurations with different numbers of turns
\cite{KulicSpools,Mihardja2006,Sudhanshu2011}.
This phenomenon is illustrated
in Fig.~\ref{LANDSCAPE}, where the final two terms of EQ. \ref{GIBBS_ELASTIC} are represented using
a grey scale, with white corresponding to the highest free energy and black the lowest.
It is clear from this figure that the periodic terms in EQ. \ref{GIBBS_ELASTIC} yield low free energies for
$(\alpha, \beta)$ near $(2 \pi,0)$, $(\pi, \pi)$, and $(0,0)$, corresponding to 2, 1 and zero turns, respectively,
and that the path from one of these minima to the next  must cross an elastic free-energy
barrier, that involves rotating the nucleosome through values of $\beta$ far from 0 or $\pi$.

\begin{widetext}
In a pulling experiment,  the external torque applied to the nucleosome is necessarily zero.
This condition implies a relationship among $F$, $\kappa$, $\alpha$,
and $\beta$ \cite{KulicSpools}, namely
\begin{equation}
F= \frac{2 \kappa  \left(\sqrt{H^2+R^2}-P \cos\alpha \cos \beta +H \sin \beta \right) (H \cos \beta +R \cos \alpha  \sin \beta )^2}{\sqrt{H^2+R^2} (H \alpha  \cos \beta +R \sin \alpha  \sin \beta )^2 
\left(
R^2 \sin^2 \alpha + \left( H \cos \beta + R \cos \alpha \sin \beta  \right )^2
\right)}.
\label{IMPLICIT}
\end{equation}
\end{widetext}
It follows that for a given force, $F$, and winding angle, $2 \alpha$, the rotation angle, $\beta$, is
prescribed by Eq. \ref{IMPLICIT}.
Although an algebraic solution for $\beta$ is not possible, it is nevertheless straightforward to
find the solutions for $\beta$ from Eq. \ref{IMPLICIT}
numerically using Mathematica (Wolfram Research, Urbana, IL).
The solution for $\beta$, satisfying EQ. \ref{IMPLICIT},  corresponding to the lowest free energy path 
in the $\alpha \beta$-plane is shown in Fig.~\ref{LANDSCAPE}. 
Evaluating Eq. \ref{GIBBS_ELASTIC},  using  $\beta$ given by Eq. \ref{IMPLICIT} leads to a
one-dimensional free energy landscape, that is a function of the winding angle alone.
According to Ref. \onlinecite{Luger1997},
126 base pairs (125 base pair spacings, henceforth 125 bp)
constitute 1.65 superhelical turns.
It follows that there are
75.8 bp  per turn.
In the remainder of this paper, we have chosen to measure the winding angle of the DNA about the
nucleosome in units of base pairs, assuming there are 75.8 bp per superhelical turn.
To convert from  $\alpha$  in radians to the winding angle in base pairs,
it is simply necessary to multiply by
$75.8 /\pi$.

The one-dimensional elastic free energy
landscape  for nucleosome winding/unwinding, is
plotted versus winding angle for several values of the force in Fig.~\ref{LANDSCAPE2}.
At each force, the free energy increases super-linearly
with increasing winding angle until a local maximum is reached,
corresponding to the elastic free energy barrier.
Both the value and the location of this maximum increase with force.
Beyond the local maximum, the free energy
varies only weakly with further increase in the winding angle
up to and including values of the winding angle
corresponding to state 1 ($\sim75$~bp).
The existence of a free energy barrier, even at large forces,
implies that state 1 is mechanically trapped in the low temperature limit.
 \begin{figure}[b!]
\begin{center}
{\includegraphics[width=3.4in,keepaspectratio=true]{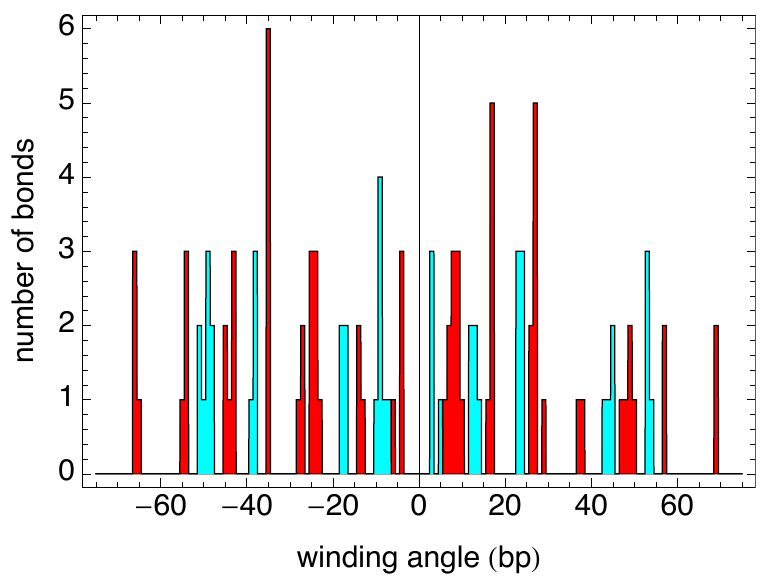}}
\end{center}
\caption{
(Color online) Histogram of the number of histone-DNA bonds versus winding angle,
measured in base pairs from the dyad axis,
recognized by {\tt NUCPLOT} (using its default parameters)
for a nucleosome assembled with a palindromic version of
 601 DNA (PDB accession code 3UT9) \cite{Chua2012}.
 Shown darker (red) and lighter (cyan)
 are the bonds between histones and DNA strand I and
 DNA strand J, respectively.
 It is important to emphasize that this is the only figure in this paper in which the
 winding angle is zero at the dyad.
 In every other figure, zero winding angle corresponds to the point
 at which no DNA is wound around the nucleosome.
 }
\label{NUCPLOT}
\end{figure}

It is important to emphasize that the elastic energy represented by EQ. \ref{GIBBS_ELASTIC}
is approximate.
In particular, the x-ray crystallographically-determined structure of the nucleosome
shows significant kinking
\cite{Luger1997,Davey2002,Chua2012}.
In addition,  recent atomic force microscopy measurements indicate that
the elastic energy of highly bent DNA, such as is realized in the nucleosome,
is lower than expected on the basis of linear
elasticity  \cite{Wiggins2006}.


 \begin{figure}[b!]
\begin{center}
{\includegraphics[width=3.4in,keepaspectratio=true]{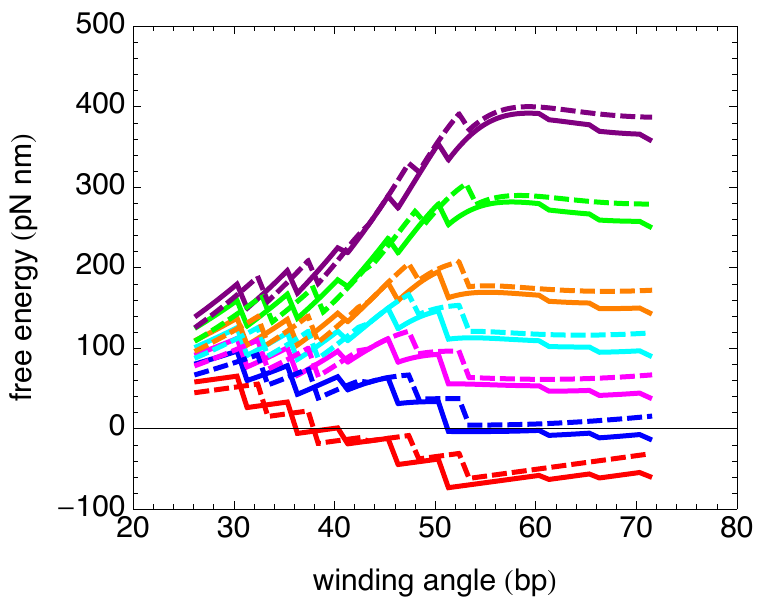}}
\end{center}
\caption{(Color online)
Model free energy landscapes for the nucleosome inner turn,
including the elastic free energy and the free energy of localized
histone-DNA binding at 5~bp intervals, at several forces.
Each step down corresponds to the disruption of one set of histone-DNA bonds.
The figure actually illustrates a binding free energy change that occurs over a small range
of winding angles (1~bp), corresponding to histone-DNA bonds distributed over this
range.
The solid curves correspond to the best fit parameters for nucleosomes
containing H4
at forces, from bottom to top, of
0, 2, 4, 6, 8, 12, and 16~pN, respectively.
The dashed curves correspond to the best fit parameters for nucleosomes
containing H4-R45H
at forces, from bottom to top, of 0, 2, 4, 6, 8, 12, and 16~pN, respectively.
}
\label{LANDSCAPE3a}
\end{figure}

\subsection{Specific histone-DNA binding}
In addition to the elastic free energy specified in EQ. \ref{GIBBS_ELASTIC},
we must also include histone-DNA binding,
which stabilizes the nucleosome.
As noted above, 
the 
measurements of
Ref. \onlinecite{Hall2009} indicate that the histone-DNA interaction
shows an approximate 5~bp periodicity, even though the
pitch of the DNA double helix is about 10~bp.
To explore histone-DNA interactions within the context of the
 crystallographically-determined nucleosome structure,
 we have used the program {\tt NUCPLOT} \cite{NUCPLOT},
 which automatically identifies protein-DNA interactions, using as input
 the Protein Data Bank (PDB) file for the protein-DNA complex of interest, {\em e.g.} the nucleosome.
Specifically,  we used {\tt NUCPLOT}, using the default criteria, to identify histone-DNA bonds
within the structure of a nucleosome assembled with a modified, palindromic version of
601 DNA, whose structure was recently determined (PDB accession code 3UT9) \cite{Chua2012}.
 A histogram of the number of bonds for each DNA strand is shown
 in Fig.~\ref{NUCPLOT}.
 Bonds between  histones and DNA in strands I and J are shown dark (red) and light (cyan),
 respectively.
 Within each strand, there is an approximate 10 bp periodicity,  but the bonds
from strand I and from strand J are staggered relative to each other,
leading to an overall approximate 5~bp periodicity.
Three regions of especially strong bonding, as suggested in Ref. \cite{Hall2009}, are
not  apparent from Fig.~\ref{NUCPLOT}.
However, Fig.~\ref{NUCPLOT} gives
the number of bonds only, and does not factor in
bond strength at all, which may be considerably different for different bonds.
What Fig.~\ref{NUCPLOT} does make clear is that the approximate 5~bp periodicity suggested
in Ref.~\cite{Hall2009} may be understood on the basis of the structure of the nucleosome.
%
%

What then is the histone-DNA free energy landscape in the presence of histone-DNA interactions?
If there is a strong localized histone-DNA bond at a particular value of the winding angle,
then the free energy will decrease in
a stepwise fashion when the winding angle is increased beyond that value,
corresponding to formation of the bond in question \cite{Forties2011}.
It follows that, if there are strong localized histone-DNA bonds spaced by 5~bp, the
histone-DNA free energy landscape as a function of winding angle will show a
corresponding sequence of stepwise decreases,
as the winding angle is increased, each step separated from its neighbors by 5~bp.
Examples of such a simplified free energy landscape at several forces,
incorporating both the elastic free energy of the DNA superhelix
under tension and the effect of histone-DNA binding in a simplified fashion, are
illustrated in Fig.~\ref{LANDSCAPE3a}.
As a result of the steps,
the net free energy landscape shows multiple minima and maxima as a function of
winding angle.
Importantly, which local free energy maximum corresponds to the global free energy maximum -- {\em i.e.}
to  the
transition state --  changes as a function of force.

\subsection{Transition rates across a piecewise linear free energy landscape}
To calculate the unwinding and rewinding rates across free energy landscapes  of
the sort shown in
Fig.~\ref{LANDSCAPE3a},
we use the ``flux over population method'' \cite{Hanggi},
which exploits the observation  that the rate in question
is equal to the normalized steady-state flux into the final state for an adsorbing boundary condition.

First, we consider a discrete model, in which each
state along the reaction pathway is labelled by an integer $m$,  with $m=0$ corresponding to the final state,
from which there are no transitions back.
Then, the transition rate is equal to the normalized steady-state flux into state 0 for
$p_0=0$,
namely
\begin{equation}
k_{n \rightarrow 0}= 
 p_1 k_{1 \rightarrow 0} .
\label{FLUX}
\end{equation}
where $p_1$ is the probability that the system is in state 1, and $k_{1\rightarrow 0}$
is the rate of transitions from state 1 to state 0.
The quantity, $k_{n\rightarrow 0}$,  is equal to the inverse of the mean first passage time (MFPT)
from state $n$ to state 0 \cite{Redner}.
The probabilities ($p_1$, $p_2$, {\em etc.})  are specified
in terms of the transition rates between neighboring states, via the steady-state master equations: 
\begin{equation}
0=  -(k_{1\rightarrow 0} +k_{1 \rightarrow 2})p_1+k_{2 \rightarrow1}p_2,
\label{EQ555}
\end{equation}
\begin{equation}
0= k_{1\rightarrow 2}p_1-(k_{2 \rightarrow 1} +k_{2 \rightarrow 3})p_2+k_{3 \rightarrow 2}p_3,
\label{EQ666}
\end{equation}
{\em etc.}
and are normalized:
$p_1 + p_2 + ... + p_n =1$.
It is then straightforward to derive a recursion relation for $k_{n \rightarrow 0}$,
namely
\begin{widetext}
\begin{equation}
\frac{1}{k_{n \rightarrow 0}} = \frac{1}{k_{n\rightarrow n-1}} +
\left ( 1+ \frac{k_{n-1\rightarrow n}}{k_{n\rightarrow n-1}} \right )
\frac{1}{k_{n-1 \rightarrow 0}} -
\frac{k_{n-1\rightarrow n}}{k_{n\rightarrow n-1}} \frac{1}{k_{n-2 \rightarrow 0}},
\label{RECURSION1}
\end{equation}
where we take $1/k_{n-2 \rightarrow 0}=0$ for $n=2$.
\end{widetext}


EQ. \ref{RECURSION1} is  useful when the individual  transition rates are
known. However, if the free energy landscape is a function
of the reaction coordinate ({\em i.e.} the winding angle, as in Fig.~\ref{LANDSCAPE3a}),
it is preferable to employ a continuum description,
which permits the transition rates to be determined in terms of parameters that
describe the free energy landscape.
In the case of a piecewise linear free energy landscape \cite{Privman1991,Palyulin2012},
it is possible to derive an analogous recursion formula to EQ. \ref{RECURSION1}, as follows.
First, we label the locations of the cusps between neighboring linear regions by an integer $m$, so that the
locations of the cusps are $L_m$. The difference in free energy between
location $L_{m+1}$ (final state) and location $L_m$ (initial state), we designate
as $\Delta G_{m \rightarrow m+1}$
Then, similarly to EQ. \ref{FLUX},
the transition rate across a piecewise linear free energy landscape from $x=L_n$ to $x=0$
is equal to the
particle flux at $x=0$ for adsorbing boundary conditions at $x=0$, namely
\begin{equation}
k_{n \rightarrow 0}  = -D  \frac{dp(0)}{dx},
\end{equation}
where
$p(x)$ is the appropriately normalized probability density, that solves
the steady-state Smoluchowski equation,
subject to the boundary conditions that $p(0)=0$ and that $p(x)$ is continuous at the
boundaries of each piecewise linear region.
(We assume that the free energy is continuous).
In this case, we can show that
\begin{widetext}
\begin{equation}
\frac{1}{k_{n \rightarrow 0}} = \sigma_n+ \frac{1}{D} \Sigma_{p=1}^n
(L_{p-1}-L_p)^2 \frac{
e^{\Delta G_{p\rightarrow p-1}/(k_B T)} - 1 - \Delta G_{p \rightarrow p-1}/(k_B T)
}{
\Delta G_{p \rightarrow p-1}^2/(k_B T)^2
}
\label{EQ13}
\end{equation}
with
\begin{equation}
\sigma_n =  \rho_n
+ \left (1+
K_n
\right )
\sigma_{n-1}
-
K_n
\sigma_{n-2},
\label{EQ14}
\end{equation}
\begin{equation}
K_n = \frac{(-1+e^{\Delta G_{n \rightarrow n-1}/(k_B T)})(L_n-L_{n-1}) /\Delta G_{n \rightarrow n-1}
}
{
(1 - e^{-\Delta G_{n-1 \rightarrow n-2}/(k_B T)})(L_{n-1}-L_{n-2}))/ \Delta G_{n-1\rightarrow n-2}
},
\label{EQ15}
\end{equation}
and
\begin{equation}
\rho_n=
- \frac{ (L_n-L_{n-1})(L_{n-1}-L_{n-2})
(-1+e^{\Delta G_{n-1 \rightarrow n-2}/(k_BT)}) (-1+e^{\Delta G_{n \rightarrow n-1}/(k_BT)})
}
{
D\Delta G_{n \rightarrow n-1}\Delta  G_{n-1 \rightarrow n-2}/(k_BT)^2
} .
\label{EQ16}
\end{equation}
\end{widetext}
Ordinarily, the second term on the right-hand-side of EQ. \ref{EQ13} is smaller than the first term.
Therefore, it may often be a good approximation to neglect the second term
with the result that
\begin{equation}
\frac{1}{k_{n \rightarrow 0}}  \simeq  \rho_n
+ \left (1+
K_n
\right )
\frac{1}{k_{n-1 \rightarrow 0}}
-
K_n
\frac{1}{k_{n-2 \rightarrow 0}},
\label{EQ14a}
\end{equation}
which is formally the same as EQ. \ref{RECURSION1},
provided we identify $K_n={k_{n-1 \rightarrow n}}/{k_{n \rightarrow n-1}}$,
and $\rho_n = {1}/{k_{n \rightarrow n-1}}$.
It is also possible to give an explicit expression for $\sigma_n$:
\begin{widetext}
\begin{equation}
\sigma_n=
\frac{1}{D}
\Sigma_{p=2}^{n}
\Sigma_{q=1}^{p-1}
\frac{ (L_{p-1}-L_{p})(L_{q-1}-L_{q})
(-1+e^{\Delta G_{p \rightarrow p-1}/(k_BT)})
e^{\Sigma_{m=q}^{p-2} \Delta G_{m+1 \rightarrow m}/(k_BT)} (-1+e^{\Delta G_{q \rightarrow q-1}/(k_BT)})
}
{
\Delta G_{p \rightarrow p-1}\Delta  G_{q \rightarrow q-1}/(k_BT)^2
} .
\label{EQ17}
\end{equation}
\end{widetext}
Inspection of EQ. \ref{EQ17} reveals that the MFPT for a landscape
with multiple minima and maxima is dominated by the term corresponding to the
largest free energy barrier,
as expected on the basis of the Arrhenius equation for
the transition rate ($k$) across a free energy barrier, $\Delta G^\dagger$, namely
\begin{equation}
k = k_0 e^{-\Delta G^\dagger/(k_BT)},
\label{ARR}
\end{equation}
where $k_0$ is the rate for zero barrier height ($\Delta G^\dagger =0$).
However, the factors of $\Delta G_{p \rightarrow p-1} /(k_B T)$ in the denominator of
the second term on the right-hand-side of
EQ. \ref{EQ13} and in the denominator of the right hand side of EQ. \ref{EQ17}  represent significant corrections to the Arrhenius form.
Although EQ. \ref{ARR} is widely-used, it represents an even cruder approximation than
EQ. \ref{EQ13}.

\subsection{Modelling the nucleosome unwinding and rewinding rates}
Although derived for a piecewise linear free energy landscape,
EQ. \ref{EQ13} through EQ. \ref{EQ16}
express the transition rates
solely in terms of free energy differences and reaction
coordinate differences.
Therefore,
these equations may be readily applied to a general free energy landscape.
This approach is equivalent to
approximating the landscape in question as a piecewise linear  landscape,
and is
the approach we have taken to model the winding and unwinding rates
for the nucleosome inner turn.
Specifically, to create a minimal model free energy landscape for the nucleosome inner turn at each force,
we add together the elastic energy given by EQ. \ref{GIBBS_ELASTIC}
and a set of linear steps,
one for each histone-DNA binding location.
For the step near base pair $m$, we take the free energy to decrease linearly
from base pair $m + \delta - 1/2$ to base pair
$m +\delta +1/2$, where $m$ is a positive integer and
$\delta$ specifies the registration between the zero
of the winding angle and
the zero of the 5 bp periodicity.

Fig.~\ref{LANDSCAPE3a} shows examples of such a model landscape.
In the following, it is convenient to denote the location of base pair $m + \delta - 1/2$
as location $m$ and the location of base pair $m + \delta + 1/2$ as location $m.5$,
and to denote the change in binding free energy from initial location $m$
to final location $m.5$ as $\Delta G_m$.
To apply EQ. \ref{EQ13} through EQ. \ref{EQ16},
we identify successive locations appearing in these equations
with locations $25.5$, $30$, $30.5$, $35$, $35.5$ {\em etc.}
Thus, we calculate the unwinding and rewinding rates across the model free energy landscape.

To fit the model rates so-obtained to the data shown in Fig.~\ref{RATES},
we are lead to introduce the following possible fitting parameters: $\delta$, which
specifies the registration between the zero of the winding angle and
 the 5 bp periodicity;  $D$, which is an effective rotational
diffusion coefficient;
and a set of histone-DNA binding energies ($\Delta G_m$).
To limit the number of fitting parameters to as few as possible,
we held fixed the DNA persistence length ($L_P=42$~nm),
the spacing between histone-DNA binding sites (5~bp),
and the width of each histone-DNA binding site (1~bp).

In both state 1 and  state 0, the number of base pairs of DNA in
contact with the histone octamer is not precisely known.
However, to numerically calculate the transition rates between state 1 to state 0,
it is necessary to pick definite starting and ending winding
angles.  For the purposes of our calculation of the transition
from state 1 (0) to state 0 (1), therefore,
we assume a starting winding angle corresponding to base pair number 70 (25) and
an ending winding angle corresponding to base pair number 25 (70),
which correspond to the free energy landscapes shown in Fig. ~\ref{LANDSCAPE3a}.
In fact, EQs. \ref{EQ13} through \ref{EQ17} show that the transition rate
is a sum of terms, and that the rate is determined by the largest
free energy barrier encountered, {\em i.e.} the largest term in EQ.~\ref{EQ17}.
It follows therefore that even if the starting and ending base pairs
are actually larger and smaller, respectively, than 70 and 25, provided there
is not a significant contribution to the free energy barrier from base pairs
between the actual starting and ending base pairs and base pairs 70 and 25, respectively,
the calculated model rates will be essentially unchanged.
Our assumption that locations 70.5 and 25.5 are reasonable end points may be
 justified {\em  a posteri} on the basis of the
success of the fits, we achieve.
The histone-DNA binding energies within the included region,
which are therefore {\em possible} fitting parameters, are:
$\Delta G_{30}$,
$\Delta G_{35}$,
$\Delta G_{40}$,
$\Delta G_{45}$,
$\Delta G_{50}$,
$\Delta G_{55}$,
$\Delta G_{60}$,
$\Delta G_{65}$,
and $\Delta G_{70}$.
We assume that each of these binding energies is negative, corresponding to binding.

Irrespective of the values of these fitting parameters, it is not possible to achieve
satisfactory agreement between our model and the
two experimental data points obtained at high forces ($F > 14$~pN) for
nucleosomes containing H4.
Therefore, we have added to our model rate an {\em ad hoc} term,
$D \exp[ \Delta G^\ddagger+F x^\ddagger/(k_B T)]$ \cite{Bell1978},
which is negligible at low forces,
but which is able to match these two data points.
In principle, such a term describes a process that is an alternative to unwinding,
and competes with unwinding. For example, we may speculate that a high-force alternative
to unwinding is that the DNA may slide sideways off of the nucleosome.

Because the values of the rates span several decades,
in order to achieve satisfactory agreement between the measured rates and our model over the full range,
we carried out least-mean-squares fits of
the logarithm of the model rates to the logarithm of the measured rates,
using Mathematica's {\tt NonLinearModelFit} function.
We sought to limit, as far as possible, the total number of fitting parameters, while still
achieving a satisfactory description of our experimental data (Fig.~\ref{RATES}).
Therefore,
we fitted all of our rate-constant data simultaneously,
so that the unwinding and rewinding rates were both described
by the same set of binding free energies.
 Furthermore, we used the same effective rotational diffusion
 constant for unwinding and rewinding and
 for nucleosomes
 containing H4 and nucleosomes containing H4-R45H. 
 Because initial fitting revealed that the rates could be well-described
with $\Delta G_{55}=0$,
this parameter was held fixed at zero for both types of nucleosome.
Initial fitting also revealed that
the rates depend only on the sum
$\Delta G_{70}+\Delta G_{65}+\Delta G_{60}$.
Therefore to
further restrict the number of fitting parameters,
we fixed $\Delta G_{70}=\Delta G_{65}=\Delta G_{60}$,
corresponding to the maximum likelihood, when only the sum is determined.
In the case of nucleosomes containing H4-R45H,
the rates could be well-described with $\Delta G_{70}=\Delta G_{65}=\Delta G_{60}=0$,
and so these parameters were fixed equal to zero in this case.
 Finally, we
fixed $\Delta G_{30}$ to be equal to $\Delta G_{35}$ and constrained these parameters
to the values that ensure that the free energy of a nucleosome
with H4 at $F=F^*=6.0$~pN is the same at locations 25.5 and 70.5,
and that the free energy of a nucleosome containing H4-R45H at $F=F^*=5.0$~pN is the same at
locations 25.5 and 70.5, consistent with the extrapolation of our experimental measurements.
The resultant best fits are shown as the solid and dashed lines in Fig.~\ref{RATES}, including
the {\em ad hoc} term for the unwinding of nucleosomes containing H4.
Evidently, the model rates
provide a good description of our measurements.
The corresponding best fit parameters are shown Table \ref{TABLE1}.
\begin{table*}
\vspace{0.2in}
\begin{tabular}{| r | r | r |}
\hline
& H4 & H4-R45H\\
\hline
$\Delta G_{30}$ & $-40.7$~pN nm ($-9.9 k_B T$)$^*$ & $-42.0$~pN nm ($-10.2 k_B T $)$^*$\\
$\Delta G_{35}$ & $-40.7$~pN nm ($-9.9k_B T$)$^*$& $-42.0$~pN nm ($-10.2 k_B T$)$^*$\\
$\Delta G_{40}$ & $-21.5 \pm 1.3$~pN nm ($-5.2\pm0.3 k_B T$)& $-5.6\pm 2.0$~pN nm ($-1.4 \pm 0.5 k_B T$)\\
$\Delta G_{45}$ & $-34.1\pm1.6$~pN nm ($-8.3\pm0.4k_B T$)& $-30.7\pm1.1$~pN nm ($-7.5\pm0.3 k_B T$)\\
$\Delta G_{50}$ & $-37.2\pm2.6$~pN nm ($-9.1\pm0.6 k_B T$)& $-32.6\pm2.2$~pN nm ($-7.9\pm0.5 k_B T$)\\
$\Delta G_{55}$ &0.0~pN nm$^{**}$& $0.0$~pN nm$^{**}$\\
$\Delta G_{60}$ & $-6.7\pm0.5$~pN nm ($-1.6\pm0.1 k_B T$)& 0.0~pN nm$^{**}$\\
$\Delta G_{65}$ & $-6.7$~pN nm ($-1.6k_B T$) $^{***}$ & 0.0~pN nm$^{**}$\\
$\Delta G_{70}$ & $-6.7$~pN nm ($-1.6k_B T$) $^{***}$ & 0.0~pN nm$^{**}$\\
$\delta$ & $0.82 \pm 0.37$~bp & $2.88 \pm 0.33$~bp\\
$D$ & $5500\pm990$~bp$^2$s$^{-1}$$^{****}$ & $5550\pm990$~bp$^2$s$^{-1}$$^{****}$\\
$\Delta G^\ddagger$ & $-128\pm25$~pN nm ($32.9\pm7.3 k_B T$) & N.A.\\
$x^\ddagger$ & $8.3\pm 1.7$~nm& N.A. \\
\hline
\end{tabular}
\caption{Best fit parameters that yield the model rates (lines) in Fig.~\ref{RATES}.
$^*$Parameters constrained to ensure that the unwinding and rewinding rates are equal at $F^*$
and equal to each other.
$^{**}$ Parameters fixed at zero.
$^{***}$ Fixed equal to $\Delta G_{60}$.
$^{****}$Rates for nucleosomes containing H4 and nucleosomes containing H4-R45H
were fit simultaneously using the same value of $D$ for both data sets.
The quoted errors correspond
to the standard errors produced by Mathematica's {\tt NonLinearModelFit}
via the {\tt ParameterErrors} property.
}
\label{TABLE1}
\end{table*}

\begin{figure}[b!]
   \begin{center}
      {\includegraphics[width=3.2in,keepaspectratio=true]{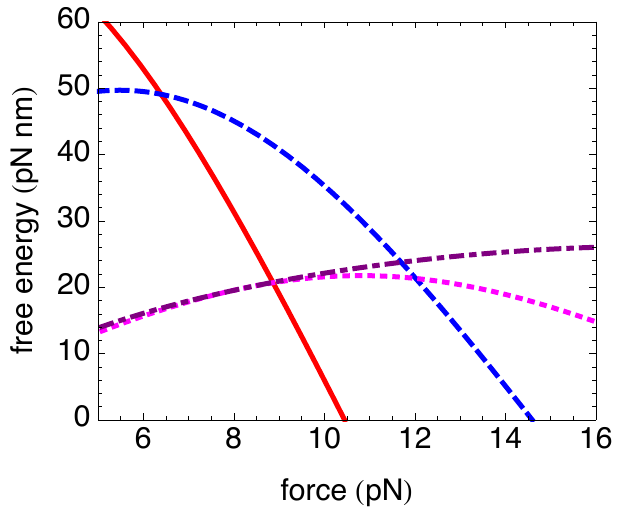}}
\caption{(Color online)
Free energy differences as a function of force,
between location 70.5 and location 45 (solid line),
location 70.5 and location 50 (dashed line),
location 70.5 and location 55 (dotted line),
and location 70.5 and the location of the elastic free energy maximum (dot-dashed line),
calculated using the best-fit parameters corresponding to nucleosomes containing H4.
The largest of these free energy differences at a given force is rate-limiting at that force.
}
\label{DECONSTRUCTION}
   \end{center}
\end{figure}

\begin{figure}[t!]
   \begin{center}
      {\includegraphics[width=3.2in,keepaspectratio=true]{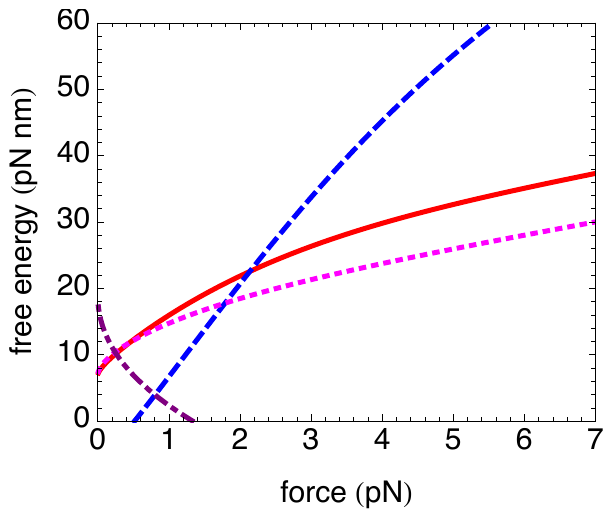}}
\caption{(Color online)
Free energy differences as a function of force between location 35.5 and location 40 (solid line),
location 35.5 and location 45 (dashed line),
location 30.5 and location 35 (dotted line),
and location 50.5 and  location 70 (dot-dashed line),
calculated using the best-fit parameters corresponding to nucleosomes containing H4.
The largest of these free energy differences at a given force is rate-limiting at that force.
}
\label{DECONSTRUCTION2}
   \end{center}
\end{figure}

\section{Discusion}

\subsection{Mapping landscape to histone-DNA binding sites}
It is important to realize that how to relate the binding locations across the free
energy landscape to specific positions within the nucleosome is not straightforward in
the absence of additional
information or assumptions. This is
because each  5 bp of DNA between successive binding locations
can detach from either side of the histone octamer, leading to the same reduction in the winding angle.

\subsection{Mapping landscape to rates}
To understand which features
of our model free energy landscape determine the unwinding and rewinding rates and
how well each of the fitting parameters is determined,
it is instructive to plot the difference in free energy between the starting location
(location 70.5 for unwinding and location 25.5 for rewinding)
and various intermediate locations on the
unwinding or the rewinding pathway.
The largest of such free energy differences
are shown using the best fit parameters for nucleosomes containing H4 in
 Fig.~\ref{DECONSTRUCTION} (unwinding) and Fig.~\ref{DECONSTRUCTION2} (rewinding).
 Each of these curves represents a  free energy barrier to unwinding.
We may expect the largest free energy barrier to unwinding at a particular force
to be rate limiting at that force.
 Clearly, for forces up to about 6.5~pN, the rate-limiting barrier to unwinding is at location 45;
 for forces between about 6.5 and 11.5~pN, the rate-limiting barrier to unwinding is at location 50;
 for forces above about 11.5~pN, the rate-limiting barrier to unwinding is at the location of the elastic
 free energy maximum.
 
 In the context of our free energy landscape model, the maximum barrier height for unwinding
 at the elastic free energy maximum depends on $\Delta G_{70}+\Delta G_{65}+\Delta G_{60}$;
at location 50, it depends additionally on $\Delta G_{50}$;
and, at location 45, it depends additionally on $\Delta G_{45}$.
 Therefore, we may expect
 $\Delta G_{70}+\Delta G_{65}+\Delta G_{60}$,
 $\Delta G_{50}$, and $\Delta G_{45}$
to be well-determined by fitting the unwinding rates.
For a given value of  their sum,
maximum likelihood corresponds to $\Delta G_{70}=\Delta G_{65}=\Delta G_{60}$.
Hence, we fixed $\Delta G_{70}=\Delta G_{65}=\Delta G_{60}$ for fitting.
 Importantly, for forces above 11.5~pN, the barrier to unwinding
 actually increases slightly with increasing force.
 This feature of the model naturally accounts for the
experimental observation  that  the
unwinding rate is nearly force-independent within the force range between 11.5
and 14~pN.
Thus, we assert that this feature of our data is especially
convincing evidence that a correct description indeed
involves the elastic free energy barrier
 in an essential fashion.
 The corresponding plot for nucleosomes containing H4-R45H (not shown)
 reveals that  within the range of forces studied experimentally
 the rate-limiting barrier is either at location 45 or at location 50.
This is a result of $\Delta G_{70}=\Delta G_{65}=\Delta G_{60}=0$
for nucleosomes containing H4-R45H.
 
For rewinding,  Fig.~\ref{DECONSTRUCTION2} reveals that
for forces between about 0.3 and 2~pN,  the free energy barrier to rewinding
corresponds to the force-dependent elastic energy involved in rewinding from
location 35.5 to location 40.
The height of this barrier is independent of any of our binding energy parameters,
except insofar as they partition the free energy landscape so that this is the highest barrier.
For forces between about 2 and 6~pN, the free energy barrier for
rewinding corresponds to the elastic energy involved in rewinding
from location 35.5 to location 45, which depends on 
 $\Delta G_{40}$, which we may therefore expect to be
well-determined by  fits to the rewinding rate.
The values of $\Delta G_{30}$ and $\Delta G_{35}$ are not well-determined by fitting. Rather
$\Delta G_{30}+\Delta G_{35}$ is constrained by knowing that the free energy of states 0 and 1 are
equal at $F^*$. The  $\Delta G_{30}=\Delta G_{35}$ is the maximum likelihood result, given that
only the sum is known.


\subsection{Alternative models}
We may inquire how well our rate measurements and model are able to discriminate against alternative
models of histone-DNA binding.
For example, a number of authors
have postulated a constant histone-DNA
binding free energy per bp
\cite{PhysRevLett.86.4414,KulicSpools,Mihardja2006,Sudhanshu2011},
albeit often with different values for the inner and outer turns.
Accordingly, we have calculated the force-dependent
rates expected in the case of a constant
histone-DNA binding free energy per base pair of 4.1~pNnm per bp,
or equivalently 12~pN.
We picked a value of 12~pN
in order to ensure that the free energy for a winding angle of 25~bp  is equal to the free energy for
a winding angle of 70~bp at a force of 6~pN.

\begin{figure}[t!]
\begin{center}
{\includegraphics[width=3.4in,keepaspectratio=true]{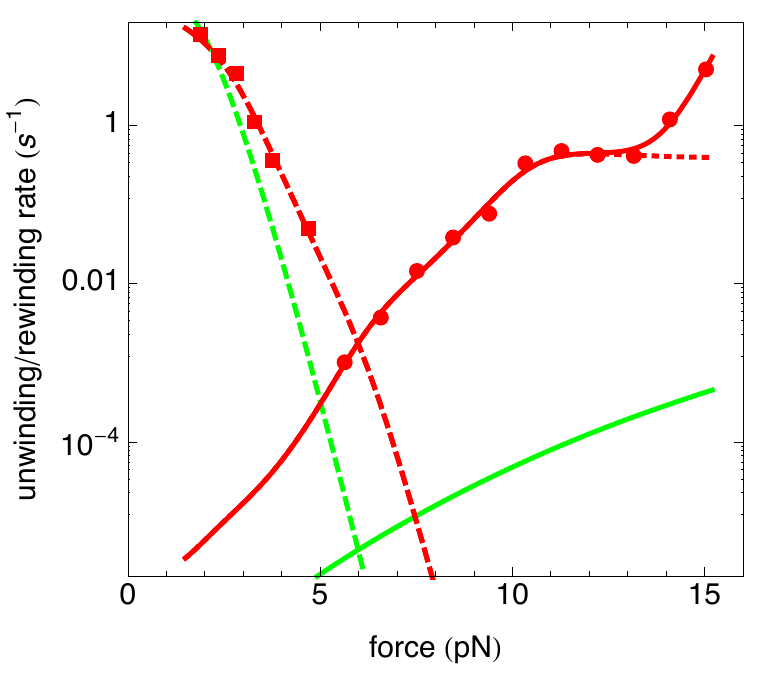}}
\end{center}
\caption{
(Color online)
Comparison between the unwinding and rewinding rates of the nucleosome
inner turn, calculated according to the local interaction model of histone-DNA
interactions,
and calculated according to a model that posits a uniform binding free energy per unit length.
The upper (red) solid and dashed  lines correspond to
the unwinding and rewinding rates, respectively, for the best-fit local
interaction model for nucleosomes containing H4
The dotted line is the local interaction model, but omitting the {\em ad hoc} term,
described in the text.
Circles and squares correspond to the measured unwinding and  rewinding rates,
respectively, for nucleosomes containing H4.
The lower solid and dashed curves correspond to the unwinding and rewinding rates, respectively, for a constant histone-DNA binding energy per bp, chosen so that at a force of 6~pN the free energies of 25 and 70~bp wound are equal.
}
\label{RATES10}
\end{figure}

\begin{figure}[t!]
\begin{center}
{\includegraphics[width=3.4in,keepaspectratio=true]{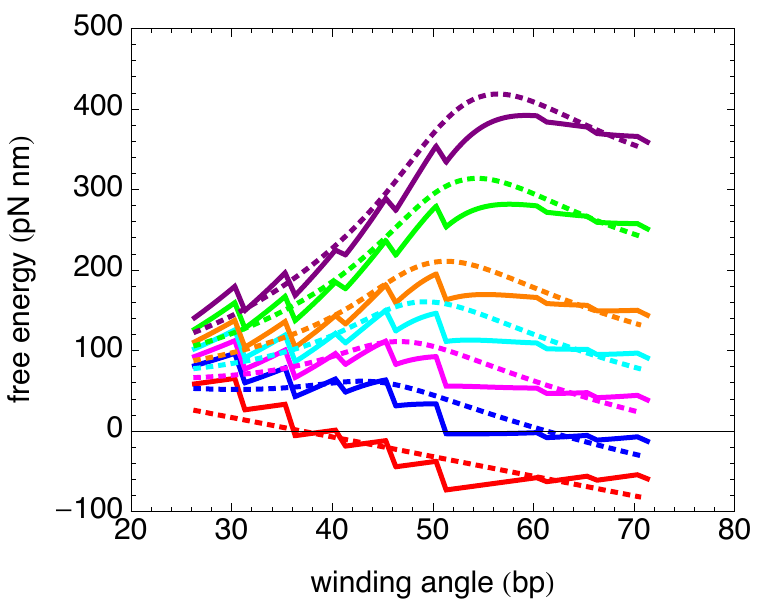}}
\end{center}
\caption{(Color online)
Model free energy landscapes for the nucleosome inner turn
at, from bottom to top,
0, 2, 4, 6, 8, 12, and 16~pN.
The solid curves reproduce those shown in Fig.~\ref{LANDSCAPE3a}, for which
each step down corresponds to the disruption of one set of histone-DNA bonds occurring over a small range
of winding angles equivalent to 1~bp.
The dotted curves correspond to a constant histone-DNA binding energy per bp (corresponding to 12~pN), chosen so that at a force of
6~pN the
free energies of 25 and 70~bp wound are equal.
}
\label{LANDSCAPE10}
\end{figure}

These calculations are compared with the experimental results
for nucleosomes containing H4 and the results of our model in Fig.~\ref{RATES10}.
The corresponding free energy landscape is shown in Fig.~\ref{LANDSCAPE10}.
The constant-binding-energy-per-bp model predicts unwinding and rewinding
rates with significantly different force-dependences
than observed experimentally and reproduced by our model.
In addition, the unwinding rate is a factor of
about 100-fold slower than observed. Thus, our data and modeling rule out such a model.
We also tested models that impose a 10~bp spacing between histone-DNA binding sites, but were unable to achieve  satisfactory fits with such a model (not shown).
 
\subsection{Effective rotational diffusion constant}
According to our fitting results,
the effective rotational diffusion coefficient of the nucleosome is 
$5500$~bp$^2$ s$^{-1}$.
In comparison, the rotational diffusion coefficient of a sphere of radius $R=4.2$~nm is
given by
\begin{equation}
D^* = \frac{k_B T}{8 \pi \eta R^3}
= 2 \times 10^{6}~{\rm s}^{-1}
\simeq 2.9 \times 10^8~{\rm bp^2 s^{-1}}
\label{D0}
\end{equation}
where $\eta=0.001$~Pas is the viscosity of the fluid in which the sphere rotates.
Therefore,  there is a factor $5 \times 10^4$ discrepancy between our measured value of $D$
and the value appropriate for free rotation of a sphere of nucleosomal radius.
How then does our best-fit value for  the effective diffusion coefficient compare to results in the literature?
Via fluorescence  experiments,
Li {\em et al.} measured the rewinding rate  for 27~bp of the nucleosome outer turn
to lie between 20 and 90~s$^{-1}$ at zero force \cite{GuLi}.
Based on our measured rotational diffusion coefficient for the nucleosome inner turn, and assuming
a flat coarse-grained free energy landscape,
we would predict a rate to rewrap 27~bp of $5500/27^2 = 8$~s$^{-1}$ close to the
experimental result of Ref. \onlinecite{GuLi}.
In fact, for a free energy landscape that is inclined downwards, as a result of histone-DNA
binding, we may expect somewhat faster rewinding and even better agreement.
Our results for $D$ also seem consistent with the extrapolations to zero force of
 the nucleosome {\em outer turn} rewinding rates
given in Refs. \cite{Mihardja2006} and \cite{Kruithof2009b} of about 100~s$^{-1}$
in both cases.

To explain  the apparent discrepancy between our result for the
rotational diffusion coefficient and  EQ. \ref{D0},
we turn to Refs. \onlinecite{Zwanzig1988} and \onlinecite{Thirumalai2003},
which
show theoretically that long-length-scale
diffusion across a free energy landscape with short-scale ``roughness''
may be described on the basis of a coarse-grained free energy landscape with an effective diffusion
coefficient that is renormalized by a factor that accounts for the short-scale roughness.
Refs. \onlinecite{Zwanzig1988} and \onlinecite{Thirumalai2003} are directly applicable here:
The free energy landscape shown in Fig.~\ref{LANDSCAPE3a} is by construction a coarse-grained
free energy, since the molecular details of the histone-DNA interaction are omitted.
It follows that the diffusion coefficient that emerges from consideration of such
a landscape is necessarily a renormalized diffusion coefficient.
Because diffusion and friction are related via the Einstein relation,
a renormalized diffusion coefficient may also be interpreted in terms of internal friction \cite{Khatri2007}.

Our best fit value of $D = 5500 \pm 990$~bp$^2$s$^{-1}$ is the effective rotational diffusion
coefficient, corresponding to the coarse-grained histone-DNA interaction free energy
of Fig.~\ref{LANDSCAPE3a}.
If we assume that the molecular-scale diffusion coefficient is $D^* = 5 \times 10^7$~rads$^2$s$^{-1}$, then
we have $D/D^*=5 \times 10^{-6}$. 
To facilitate comparisons with the results of Refs.  \onlinecite{Khatri2007} and \onlinecite{Nevo2005},
we estimate the ''roughness'' of the histone-DNA interaction at the scale of individual bonds
via  the expression for Gaussian landscape roughness, given in Ref. \onlinecite{Zwanzig1988}, namely
$\Delta G = k_B T\sqrt{ \log(D^*/D)}$. Thus, we find $\Delta G \simeq 3.5 k_B T$.
In comparison, Ref. \onlinecite{Nevo2005} determines a roughness of $5.7k_B T$ for
the forced unbinding of the GTPase Ran from the nuclear transport receptor importin-$\beta$,
and Ref. \onlinecite{Khatri2007} determines a roughness of $4k_B T$ for stretching
cellulose and dextran. Thus, our results are in-line with these studies, as well as with
Ref. \onlinecite{GuLi}.
In fact, the reaction rates in single molecule experiments are generally orders of magnitude smaller than
naive expectations. Presumably, this is because free energy landscape roughness is ubiquitous.

\subsection{Asymmetric unwinding model}
\begin{figure}[t!]
\begin{center}
{\includegraphics[width=3.2in,angle=0,keepaspectratio=true]{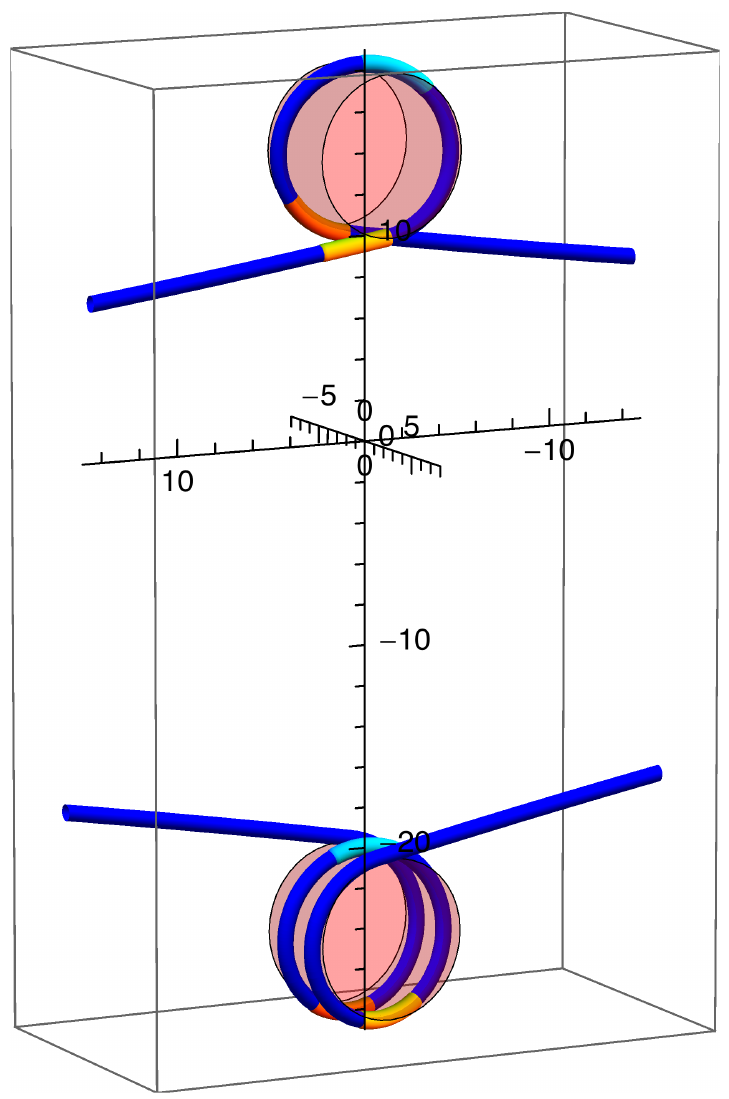}}
\end{center}
\caption{(Color online)
Elastic-rod models of nucleosomal DNA in state 1 at 8~pN, depicted as 75~bp  (top),
and in state 2  at 2~pN, depicted as 147~bp (bottom).
The three regions of DNA that form strong histone-DNA bonds
in the canonical nucleosome structure (state 2),
according to Ref. \cite{Hall2009}, are shown lighter
(orange, cyan and yellow for region 1, the dyad region/region 2,
and region 3, respectively).
In the figure, in state 2,  the DNA is depicted
symmetrically arranged about the center of nucleosome dyad (cyan).
However, in state 1, in which about 75 base pairs of DNA are wound around the histone octamer,
because regions 1 and 3 are separated by more
than about 75 base pairs (see Fig. 1 of Ref. \cite{Hall2009}),  
histone-DNA bonds in only two of the three regions of strong histone-DNA binding
-- either regions 1 and 2 or regions 2 and 3 --  can
be simultaneously satisfied.
This  is indicated in the figure by the region-3 DNA (yellow), that was bound to the histone
octamer in state 2,
being not in contact with the histone octamer, and therefore not bound, in state 1.
According to this hypothesis, symmetry about the nucleosome dyad
is not preserved in state 1, as illustrated in the figure.
The axes are marked in nanometers.
}
\label{ONETURN}
\end{figure}

Why are the unwinding and rewinding rates of nucleosomes containing H4 different from those
of nucleosomes containing H4-R45H?
To answer this question, we return to the
histone-DNA interaction map presented in Ref. \cite{Hall2009}.
In the context of the nucleosome inner turn, which contains a total of about 76~bp,
inspection of the interaction map
indicates that, if the nucleosome inner turn
were to consist of DNA that is
wound symmetrically $\pm 38$~bp from the dyad axis of the histone octamer, then the majority
 of both regions 1 and 3
would lie beyond the nucleosome inner turn.
On the other hand, if  the nucleosome inner turn is asymmetric with respect to the
dyad axis and consists of
base pair -55 to base 21, for example, relative to the dyad,
then  all or almost all of the binding energy
from both region 1 and
region 2 stabilizes the nucleosome inner turn.
In this scenario,
region 3 is involved solely in stabilizing the nucleosome outer turn.
We therefore hypothesize, as shown schematically in Fig.~\ref{ONETURN},
that the nucleosome under tension in state 1 realizes an asymmetric configuration,
because of its lower free energy compared to the symmetric configuration.
That an asymmetric partially-unwound nucleosome
configuration could be realized at zero force was suggested
previously in Ref. \cite{Polach}.

\begin{figure}[t!]
\begin{center}
{\includegraphics[width=3.2in,angle=0,keepaspectratio=true]{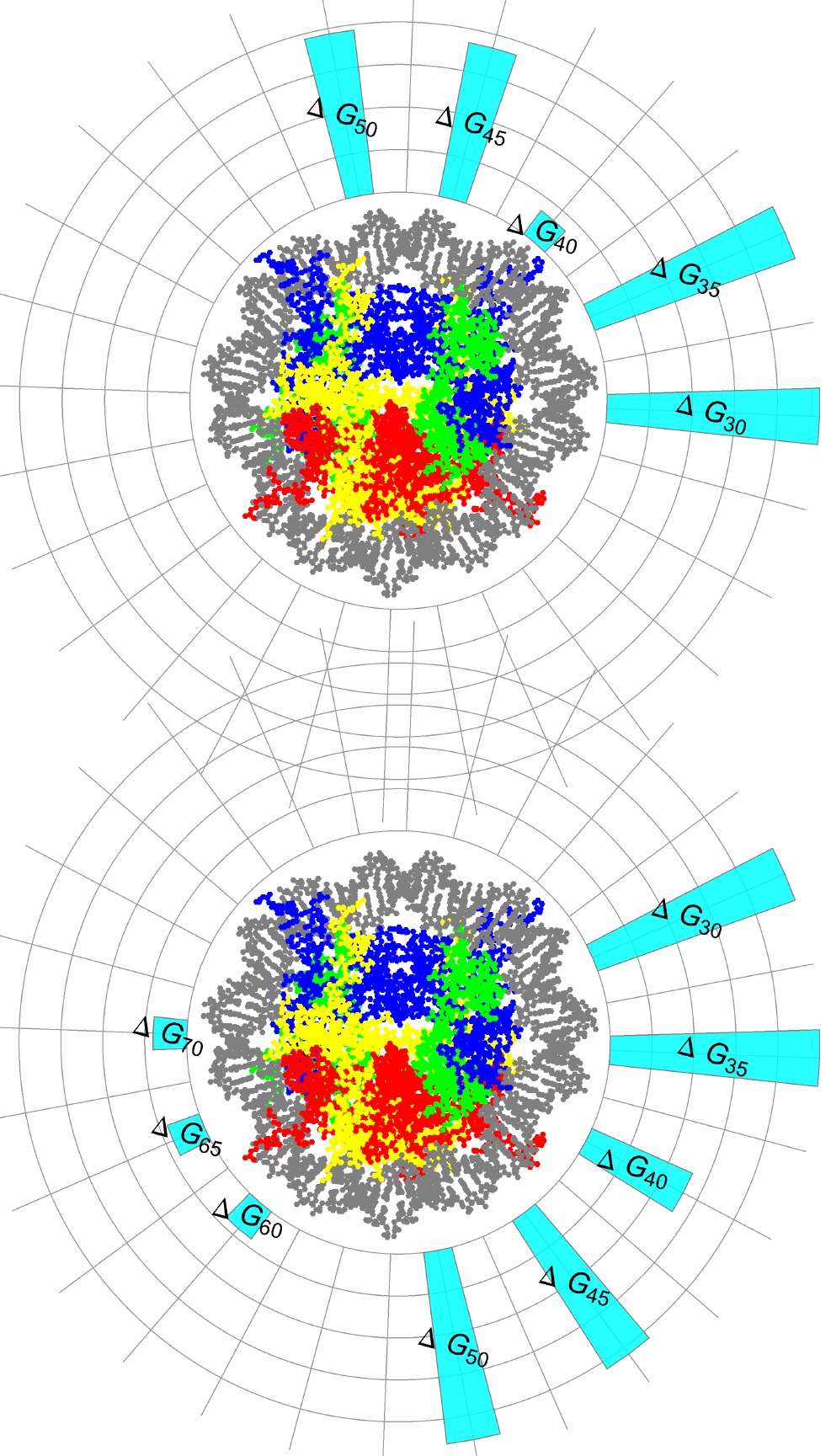}}
\end{center}
\caption{(Color online)
Hypothetical mapping of binding energies to nucleosome
locations.
Bottom: Hypothetical mapping of bond energies to nucleosome locations for nucleosomes
containing H4.
Top: Hypothetical mapping of bond energies to nucleosome
locations for nucleosomes containing H4-R45H.
The height of the bars corresponds to the binding energy at that location.
The innermost circular grid line corresponds to zero binding energy,
the next to $2 k_B T$, the next to $4 k_B T$, the next to $6k_BT$, and the outermost to $8k_B T$.
}
\label{Fig12}
\end{figure}

According to this hypothesis, in order to unwind the nucleosome inner turn, it is necessary
to break the strong histone-DNA bonds in either region 1 or region 2.
If the binding in region 1 is weaker than in region 2 \cite{Hall2009,Forties2011},
we may expect unwinding to occur via breaking region 1 bonds.
We hypothesize that this is the situation in the case of nucleosomes containing H4.
Assuming a linear relationship between
binding location and position within the nucleosome,
we may expect that the strong bonding near the dyad spans winding angles from 0 to about
25~bp, and the landscape described by our fitting parameters for
nucleosomes containing H4 corresponds to region 1.
This hypothetical situation is
illustrated in the bottom panel of Fig.~\ref{Fig12}.
By contrast, for nucleosomes containing H4-R45H, we suppose
that the binding energy in region 2 is  reduced, so that
it  is now smaller than the
binding energy in region 1. As a result, in this scenario,
unwinding occurs by breaking region 2 bonds,
and the landscape described by our fitting parameters in the
case of nucleosomes with H4-R45H corresponds to region 2.
This hypothetical situation is
illustrated in the top panel of Fig.~\ref{Fig12}.
According to this hypothesis,
the bonds disrupted  in the transition from state 1 to state 0 are different for
nucleosomes containing H4 than for nucleosomes containing H4-R45H.
Therefore, as a result,
 we may expect that the force-dependent rate for unwinding  the nucleosome
inner turn of nucleosomes containing
H4-R45H will be different than that for nucleosomes containing H4, as we observe experimentally.
Furthermore, on this basis, we {predict}  that nucleosomes containing H4 and nucleosomes containing
H4-R45H show different dissociation rates to the unbound state.
However, since unwinding and rewinding the nucleosome outer turn does not involve region 2 bonds,
we also predict that the winding and unwinding rates of these
nucleosomes' outer turn are the same  for
nucleosomes containing H4 and nucleosomes containing H4-R45H.

\section{Conclusions}
Building on the experiments presented in Ref. \cite{Mack2012a},
the principal result of this paper is that the force-dependent unwinding and rewinding rates of the nucleosome
inner turn can be explained (except at the highest forces)
on the basis of a simple, physical model that incorporates in an essential fashion both
the elastic free energy
barrier to unwinding and rewinding introduced in Ref.~\cite{KulicSpools} and elaborated in
Refs.~\cite{Mihardja2006,Sudhanshu2011}, and
localized histone-DNA binding with an approximate 5~bp periodicity,
as proposed in Ref. \cite{Hall2009}.
This analysis  provides new insight into nucleosome winding and unwinding,
the energetics of histone-DNA interactions 
and will be important for the growing numbers of simulations of nucleosome and chromatin behavior
that are appearing in the literature
\cite{Grigoryev2009,Schlick2009,Materese2009,Pono2009,Wocjan2009,PhysRevE.82.031909,Stehr2010,Ettig2011,Biswas2011,Korolev2011,Cao2011,Voltz2012,Kulaeva2012,Dobro2012}.
Beyond an improved understanding of nucleosome  unwinding and rewinding,
these results also have important implications for theories of nucleosome
sliding \cite{PhysRevLett.86.4414,PhysRevLett.99.058105,PhysRevE.79.031922,Pad2011,Biswas2012},
other processes that involve partial or complete unwinding of dsDNA from the histone octamer \cite{Saha2006}, and the mechanisms of eukaryotic gene expression more generally \cite{Boeger2003,Adkins2007,Kim2008}.

\begin{acknowledgments}
We thank J. Antonypillai, C. Cheng,  R. Ilagen, and J. Marko for valuable discussions.
This work was supported by the Raymond and Beverly Sackler Institute for Biological, Physical and Engineering Sciences and NSF PoLS 1019147. D.J.S. is the recipient of a NSF Graduate Research Fellowship.
M. K. acknowledges the support of a
NSF Postdoctoral Research Fellowship in Biology Award  DBI-1103715.

\end{acknowledgments}


%

\end{document}